\documentclass[preprint2]{aastex}

\begin{document}


\title{Common origin of two RR Lyrae populations and\\
the double red clump in the Milky Way bulge}

%% Use \author, \affil, plus the \and command to format author and affiliation 
%% information.  If done correctly the peer review system will be able to
%% automatically put the author and affiliation information from the manuscript
%% and save the corresponding author the trouble of entering it by hand.
%%
%% The \affil should be used to document primary affiliations and the
%% \altaffil should be used for secondary affiliations, titles, or email.

%% Authors with the same affiliation can be grouped in a single
\author{Young-Wook Lee and Sohee Jang}
\affil{Center for Galaxy Evolution Research and Department of Astronomy, Yonsei University, Seoul 03722, Korea; ywlee2@yonsei.ac.kr, sohee@galaxy.yonsei.ac.kr}
% \\
%2000 Florida Ave., NW, Suite 300 \\
%Washington, DC 20009-1231, USA}
%\altaffiltext{1}{ywlee2@yonsei.ac.kr}
%\altaffiltext{2}{sohee@galaxy.yonsei.ac.kr}

\begin{abstract}

A recent survey toward the Milky Way bulge has discovered two sequences of RR Lyrae stars on the period-amplitude diagram with a maximum period-shift of $\Delta log$~P $\approx$ 0.015 between the two populations. Here we show, from our synthetic horizontal-branch models, that this period-shift is most likely due to the small difference in helium abundance ($\Delta$Y = 0.012) between the first and second-generation stars (G1 and G2), as is the case in our models for the inner halo globular clusters with similar metallicity ([Fe/H] $\approx$ -1.1). We further show that the observed double red clump (RC) in the bulge is naturally reproduced when these models are extended to solar metallicity following $\Delta$Y/$\Delta$Z $\approx$ 6 for G2, as would be expected from the chemical evolution models. Therefore, the two populations of RR Lyrae stars and the double RC observed in the bulge appear to be different manifestations of the same multiple population phenomenon in the metal-poor and metal-rich regimes respectively. 
%includes a history of \aastex\ and documents the new features in the latest

\end{abstract}


\section{Introduction} \label{sec1}

Recent analysis of some thirty-thousand type ab RR Lyrae (RR$_{\rm ab}$) stars detected by OGLE survey toward the Milky Way bulge has discovered two stripes of stars in the I-band amplitude versus period diagram \citep{Pie15}. These two sequences of RR$_{\rm ab}$ stars are separated by $\Delta log$~P $\approx$ 0.015 ($\Delta$P $\approx$ 0.02 day) at low amplitudes, in the sense that ``population B" has a longer period than ``population A" at the given amplitude. \citet{Pie15} have ascribed this to the metallicity difference between the two populations ($\Delta$[Fe/H] $\approx$ 0.17), but it is unlikely from the horizontal-branch (HB) models that this metallicity difference alone can produce the observed period-shift \citep[][see Section~\ref{sec2} below]{Swe87,Lee90,Jan14}. The difference in metallicity between the two populations is also not clear as it is estimated to be negligible ($\Delta$[Fe/H]  $<$ 0.02 dex) when the difference is obtained from the modal or median values instead of the average values \citep[see Section 6 of][]{Pie15}. The origin of this phenomenon among bulge RR$_{\rm ab}$ stars is therefore yet to be understood. Since RR Lyrae stars trace old metal-poor population in the bulge, this phenomenon is potentially important to understand the early formation of the Milky Way bulge.

It is now well established that most globular clusters (GCs) in the Milky Way host multiple stellar populations with different helium and light-element abundances \citep[][and references therein]{Dan04,Lee05,Car09,Gra12}. That this effect is  responsible for the Oosterhoff dichotomy among RR$_{\rm ab}$ stars in GCs and halo fields was recently shown by \citet{Jan14} and \citet{Jan15}. According to these studies, RR$_{\rm ab}$ stars in the Oosterhoff group I GCs are mainly produced by the first generation stars (G1), while in the group II GCs, the instability strip (IS) is mostly occupied by mildly helium enhanced second generation stars (G2). Because of the small helium difference ($\Delta$Y $\approx$ 0.01), G2 in these models would be considered as a sodium-poor part of ``I" component in other studies on GCs \citep{Car09} where the subpopulations are defined by sodium and oxygen abundances. In the metal-rich regime, however, the helium difference could be significantly large in view of the chemical evolution, and in such a case, \citet{Lee15} have recently suggested that the double red clump (RC) observed in the Milky Way bulge \citep{Mcw10,Nat10} can also be reproduced by G1 and G2.\footnote{That this multiple population scenario can also reproduce the longitude dependence of the RC luminosity function at b = -8.5$\arcdeg$ \citep{Gon15} is explained by \citet{Joo16} with a composite bulge where a classical bulge component (with G1 and G2) is embedded in a bar. From the WISE mid-IR image, \citet{Nes16} report a direct detection of a faint X-shaped structure in the Milky Way bulge, but this is most likely an artifact because an ellipsoid, instead of a boxy structure, was subtracted from the bulge image \citep[D. Han \& Y.-W. Lee, in prep.; see also][]{Lop16}. Furthermore, a case against the X-shaped bulge is presented by \citet{Lop16} from the density distribution of main-sequence stars along the line of sight toward the bulge.} The purpose of this paper is to show that the two populations of RR$_{\rm ab}$ stars observed in the Milky Way bulge, at intermediate metallicity ([Fe/H] $\approx$ -1.1), are another manifestation of the same multiple population phenomenon, making the links between the metal-poor halo and the metal-rich bulge.\\
%\section{Two populations of RR Lyrae stars in the bulge} \label{sec:style}

\begin{figure}[ht!]
\figurenum{1}
\centering
\epsscale{2.0}
\plotone{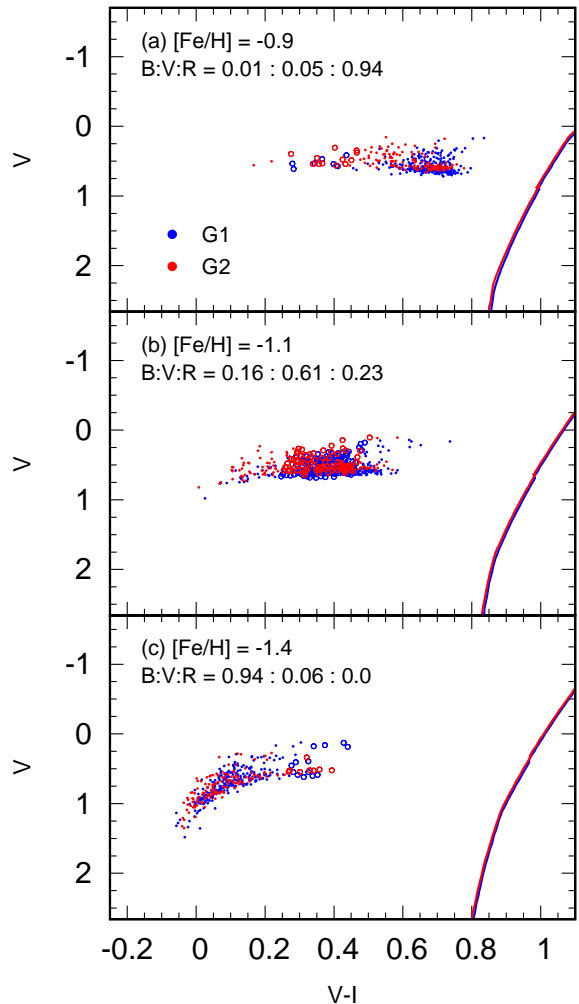}
\caption{Variation of HB morphology and corresponding change in the population of IS with metallicity for old population in the bulge. Proportions of blue HB, RR Lyrae variable, and red HB stars (B:V:R) are indicated. Note that the IS is maximally populated by RR Lyrae variables when HB type is $\sim$0.0 at [Fe/H] = -1.1. In these models, ages (13.5 and 13.0 Gyr for G1 and G2) and $\Delta$Y(G2 - G1) = 0.012 are held fixed.\label{fig1}}
\end{figure}
\section{Two populations of RR Lyrae stars in the bulge} \label{sec2}

In order to reproduce the bulge RR Lyrae stars in the multiple population paradigm, we have constructed synthetic horizontal-branch (HB) models including RR Lyrae variables following the techniques and prescriptions described in \citet{Jan15} and \citet{Lee90}. Our models are based on the most updated Yonsei-Yale (Y$^2$) HB evolutionary tracks and isochrones with enhanced helium abundance \citep{Han09,Lee15}, all constructed under the assumption that [$\alpha$/Fe] = 0.3. Following \citet{Jan15}, the \citet{Rei77} mass-loss parameter $\eta$  was employed to be 0.50, and the mass dispersion on the HB was adopted to be $\sigma_{M}$ = 0.010 M$_{\sun}$ for each subpopulation. The readers are referred to \citet{Jan15} and \citet{Joo13} for the details of our model construction.

\begin{figure}[ht!]
\vspace{-3pt}
\figurenum{2}
\centering
\epsscale{1.15}
\plotone{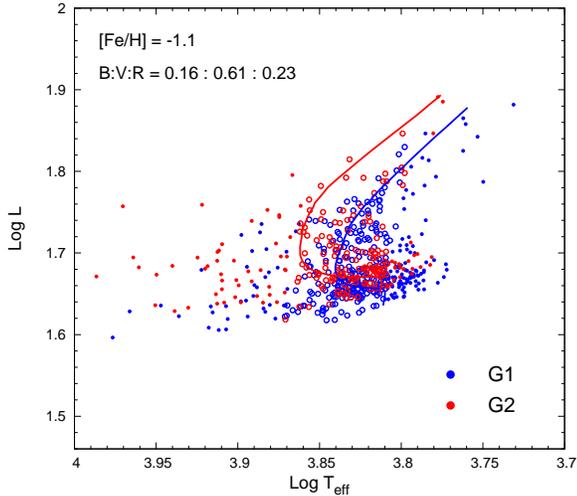}
\caption{Our synthetic HB models in the HR diagram for the bulge RR Lyrae and HB stars at the mean metallicity of bulge RR Lyrae stars ([Fe/H] = -1.1). Open circles are RR Lyrae variables, and blue and red lines are typical HB evolutionary tracks within the IS for G1 and G2, respectively. Because of a small difference in helium abundance between G2 and G1 ($\Delta$Y = 0.012), RR Lyrae stars from G2 are slightly brighter ($\Delta log$~L $\approx$ 0.02).\label{fig2}}
\end{figure}

Recent metallicity measurements of bulge RR Lyrae stars show sharply peaked metallicity distribution with an average [Fe/H] in the range of -1.25 to -1.03, depending on the method and metallicity scale \citep{Wal91,Kun08,Pie15}. In our modeling, we therefore adopt [Fe/H] = -1.1 for the RR$_{\rm ab}$ stars in the bulge. According to \citet{Lee92}, and as illustrated in Figure~\ref{fig1}, the metallicity distribution function of RR Lyrae variables is mainly dictated by HB morphology and its variation with metallicity. This implies that the HB type\footnote{The HB type is defined by the quantity, (B-R)/(B+V+R), where B, V, and R are the numbers of blue HB, RR Lyrae variable, and red HB stars, respectively.} \citep[as difined by][]{Lee94} at this peak metallicity must be $\sim$0.0, because otherwise the IS would not be maximally populated by RR Lyrae variables. In the multiple population paradigm, we found that this condition is fulfilled when we adopt model parameters similar to those required in \citet{Jan15} to reproduce the observed Oosterhoff dichotomy in the inner halo GCs. The best-fitting parameters of our models for the bulge RR$_{\rm ab}$ stars are listed in Table~\ref{tab1}. We found that this parameter combination is also required to reproduce the observed difference in period ($\Delta log$~P $\approx$ 0.015 at low amplitudes) between the two RR Lyrae populations in the bulge.

\begin{figure}[ht!]
\vspace{5.5pt}
\figurenum{3}
\centering
\epsscale{1.44}
\plotone{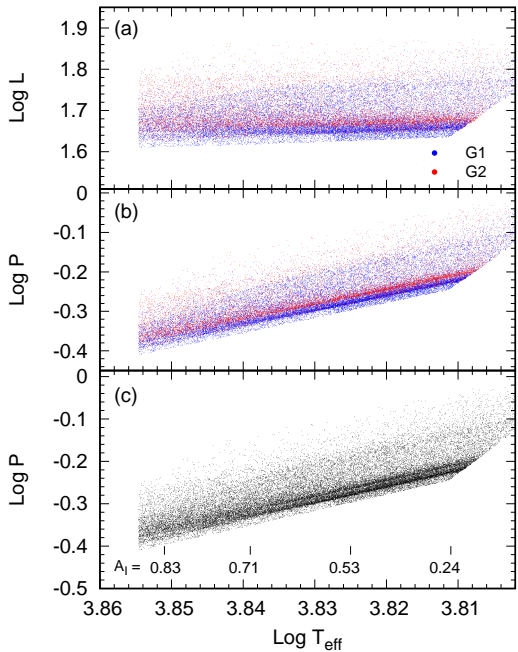}
\vspace{-15pt}
\caption{Our population models for the bulge RR$_{\rm ab}$ stars. Models in panel (a) are identical to those in Figure~\ref{fig2}, but only for the RR$_{\rm ab}$ stars with the number of stars increased to $\sim$30,000. Panels (b) and (c) show the same models in log T$_{\rm eff}$-log P diagram. Models in panel (c), plotted in the same color for both G1 and G2, can be directly compared with two populations of RR$_{\rm ab}$ stars observed in the bulge \citep[see Figure 13 of][]{Pie15}. Four tick marks in panel (c) are to indicate rough estimates of I-band amplitudes at different temperatures.\label{fig3}}
\vspace{-15pt}
\end{figure}

Figure~\ref{fig2} shows our synthetic HB model in the HR diagram at the mean metallicity of bulge RR Lyrae stars ([Fe/H] = -1.1). This model is identical to that in panel (b) of Figure~\ref{fig1}, and here we can see that the IS is populated by both G1 and G2. RR Lyrae stars from G2 are slightly brighter ($\Delta log$~L $\approx$ 0.02), which is mostly due to a small difference in helium abundance ($\Delta$Y= 0.012) between G2 and G1.  Evolutionary tracks over-plotted on this diagram explain how the clump of stars can form at the beginning of the ``blueward nose" part of the track, where the evolutionary speed is relatively slow. 
Models in Figure~\ref{fig3} are identical to those in Figure~\ref{fig2}, but the number of RR$_{\rm ab}$ stars is increased to $\sim$30,000 and the log T$_{\rm eff}$ versus log P diagram is also plotted. Panel (c) of Figure~\ref{fig3} is plotted in the same color for both G1 and G2 to compare with the two populations of RR$_{\rm ab}$ stars observed in the bulge \citep[see Figure 13 of][]{Pie15}. In the observed diagram, I band amplitude was used instead of log T$_{\rm eff}$, but since there is a good correlation between amplitude and temperature among RR Lyrae stars, our models can be compared with the observation. Four tick marks are added in panel (c) of Figure~\ref{fig3} to provide rough estimates of I-band amplitudes at different temperatures, which were calculated from the maximum points on the empirical relation between the I-band amplitude and log P of bulge RR$_{\rm ab}$ stars \citep[Figure 15 of][]{Pie15}, together with a small shift in amplitude (-0.11) to match the observed value at the fundamental blue edge.

\begin{figure}[ht!]
\vspace{11pt}
\figurenum{4}
\centering
\epsscale{1.15}
\plotone{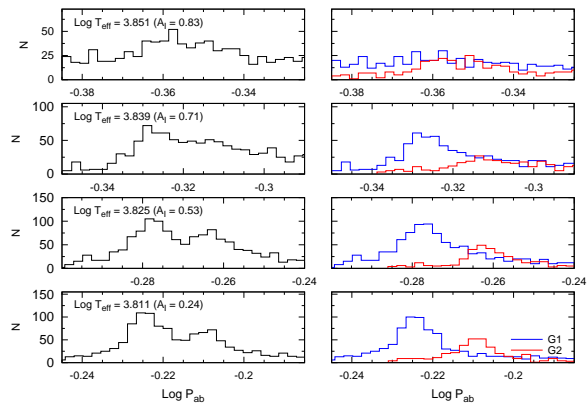}
\vspace{-15pt}
\caption{Histograms showing a period distribution of model RR$_{\rm ab}$ stars in Figure~\ref{fig3} at different temperatures. Right panels show the distributions of G1 and G2 in different colors, while the left panels show the combined distributions. Our models can reproduce the observed bimodal distribution \citep[see Figure 15 of][]{Pie15}.\label{fig4}}
\end{figure}

%\floattable
\begin{deluxetable}{cccccc}
\tablecolumns{6}
\setlength{\tabcolsep}{20pt}
\tablewidth{0pc}
%\tabletypesize{\scriptsize}

%\setlength{\tabcolsep}{18pt}

\tablenum{1}
%\center
\tablecaption{Best-fit model parameters for the bulge RR Lyrae and RC populations\label{tab1}}
\tablehead{
%\colhead{} & \multicolumn{5}{c}{Bulge}\\
\colhead{} & \multicolumn{2}{c}{Bulge RR Lyrae} & \colhead{} & \multicolumn{2}{c}{Bulge RC stars}\\
\cline{2-3}
\cline{5-6}
\colhead{Parameters} & \colhead{G1} & \colhead{G2} & \colhead{} & \colhead{G1} & \colhead{G2}
}
\startdata
{Age (Gyr)}&13.5&13.0& {}&11.0&9.0\\
{[Fe/H]\tablenotemark{a}}& -1.1 & -1.1 & {} & -0.1 & 0.1\\
{Y}& 0.235 & 0.247 & {} & 0.276 & 0.406\\
{Fraction} & 0.65 & 0.35 &{}& 0.50 & 0.50\\
{$<$M$_{\rm HB}$$>$\tablenotemark{b}} & 0.608 & 0.604 &{} &  0.814 & 0.669\\
\enddata
\tablenotetext{a}{[$\alpha$/Fe] = 0.3}
\tablenotetext{b}{Mean mass on the HB in solar units}
\vspace{-20pt}
\end{deluxetable}

\begin{figure}[ht!]
\vspace{-7pt}
\figurenum{5}
\centering
\epsscale{1.43}
\plotone{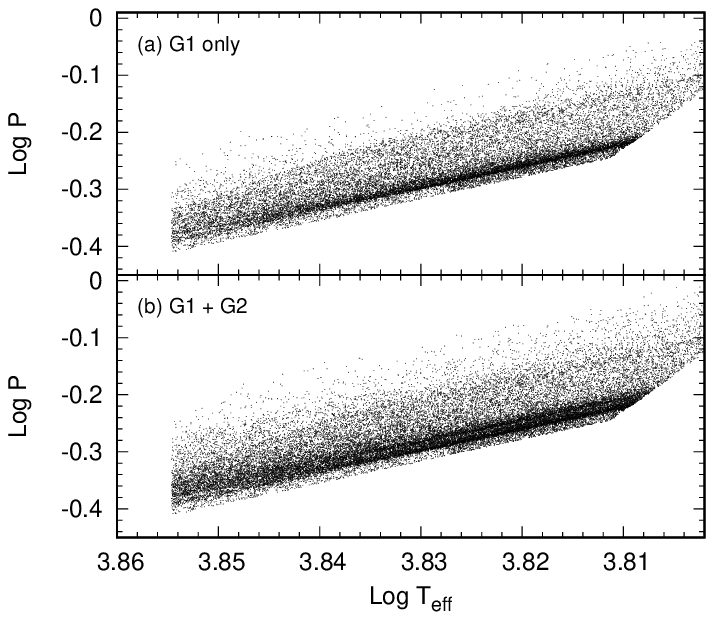}
\caption{Similar to panel (c) of Figure 3, but to compare the models with and without G2. Note that single population model in panel (a) cannot reproduce the observed two stripes of RR$_{\rm ab}$ stars.\label{fig5}}
\end{figure}

As is clear from panels (b) and (c) of Figure~\ref{fig3} (see also Figure~\ref{fig4} for the histograms), two populations of RR$_{\rm ab}$ stars are well reproduced on this diagram as two stripes of stars. As explained above, these aggregations of stars form on the ``blueward nose" part of the HB evolutionary tracks. It is clear from Figure~\ref{fig5} that the single population model (with only G1) cannot reproduce the observed two stripes of RR$_{\rm ab}$ stars. In our multiple population models, the difference in period between G2 and G1 is mostly due to the small difference in luminosity, which is in turn due to the difference in helium abundance ($\Delta$Y= 0.012).\footnote{It is well known from GC observations that stars in He-enhanced G2 also show variations in light-element (C, N, O, Na, etc.) abundances. Because of low abundances, most of these elements, excluding CNO sum, have only negligible effect on stellar models \citep[see, e.g.,][]{Van12}. Some GCs, such as NGC 1851 \citep{Yon15}, are reported to contain CNO enhanced stars, but more observations are needed to confirm the ubiquity of these stars in GCs and bulge fields. \citet{Jan14} have shown that some mild CNO enhancement ($\Delta Z_{CNO}$ $\approx$ 0.0003), in addition to helium enhancement, would be required to best reproduce the Oosterhoff dichotomy among halo GCs and RR Lyrae stars. If we had included this CNO enhancement in our models, the required $\Delta$Y would be decreased from $\sim$0.012 to $\sim$0.008 to obtain the same difference in period between the two RR Lyrae populations in the bulge.} The population ratio between G2 and G1 (G2/G1=0.54) is also well reproduced, if the observed ratio is defined from the peak positions on the histogram showing a period distribution \citep[see Figure 15 of][]{Pie15}. It is interesting to see that in both observations and our models, the separation between G1 and G2 is clearer at a lower temperature regime in the IS, while the two populations tend to overlap each other at a higher temperature regime. Additional simulations confirm that this trend is not due to the low amount of stars in the high amplitude regime. According to our models, this is most likely due to the HB tracks of G1 and G2 crossing each other at higher temperature regime within the IS (see Figure~\ref{fig2}). Therefore, within the multiple population paradigm of \citet{Jan15}, two populations of RR Lyrae stars observed in the bulge could be well reproduced by G1 (with normal He) and G2 (with enhanced He).

\begin{figure}[ht]
\vspace{0pt}
\figurenum{6}
\centering
\epsscale{1.57}
\plotone{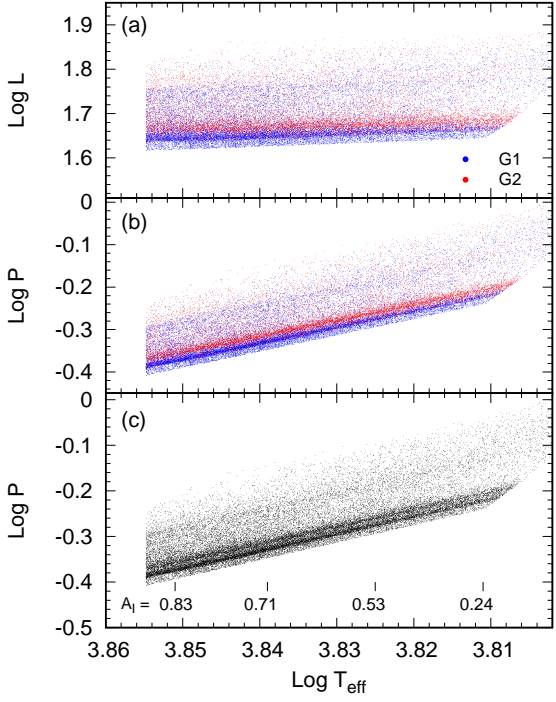}
\vspace{-10pt}
\caption{Same as Figure~\ref{fig3}, but a lower metallicity ([Fe/H] = -1.2) is adopted in the simulation so that the HB morphology is shifted to bluer color (see the text).\label{fig6}}
\vspace{35pt}
\end{figure}

\begin{figure}[ht!]
\vspace{-17pt}
\figurenum{7}
\centering
\epsscale{1.15}
\plotone{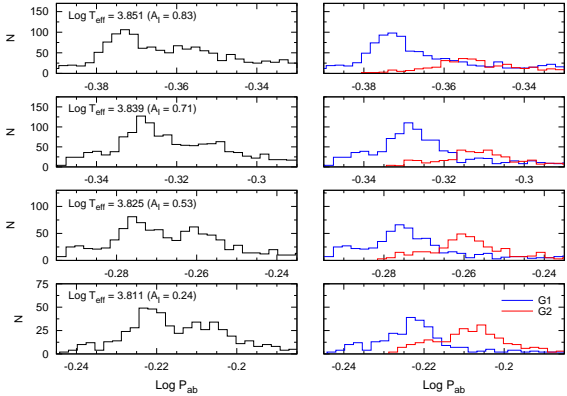}
\caption{Same as Figure~\ref{fig4}, but for a lower metallicity ([Fe/H] = -1.2) simulation in Figure~\ref{fig6}.\label{fig7}}
\vspace{23pt}
\end{figure}

An apparent drawback of our simulations in Figure~\ref{fig3} is that our models show more stars at lower temperatures (amplitudes) than at higher temperatures (amplitudes), which is contrary to what is inferred from the observed amplitude distribution \citep[see Figure 15 of][]{Pie15}. As shown in Figures~\ref{fig6} and \ref{fig7}, a better match with the observation, in terms of the distribution of stars within the IS, can be achieved by shifting the HB morphology to bluer color (HB type = +0.28), which in turn is obtained by a small decrease ($\Delta$[Fe/H] $\approx$ 0.1 dex) in metallicity. It is not clear, however, that this is a better representation of RR Lyrae stars in the bulge, because the correlation between amplitude and log T$_{\rm eff}$ is highly nonlinear \citep[see, e.g., Figure 8 of][]{Cor01}, and therefore the amplitude distribution of RR$_{\rm ab}$ stars could appear rather stretched towards a low amplitude regime. Furthermore, the completeness of RR$_{\rm ab}$ sample is most likely to be decreased at low amplitudes \citep[][]{Sos14}, which can also contribute to the bias in the distribution. In any case, two sequences of RR$_{\rm ab}$ stars are clearly reproduced in both simulations.

\begin{figure}[ht!]
\figurenum{8}
\centering
\epsscale{1.57}
\plotone{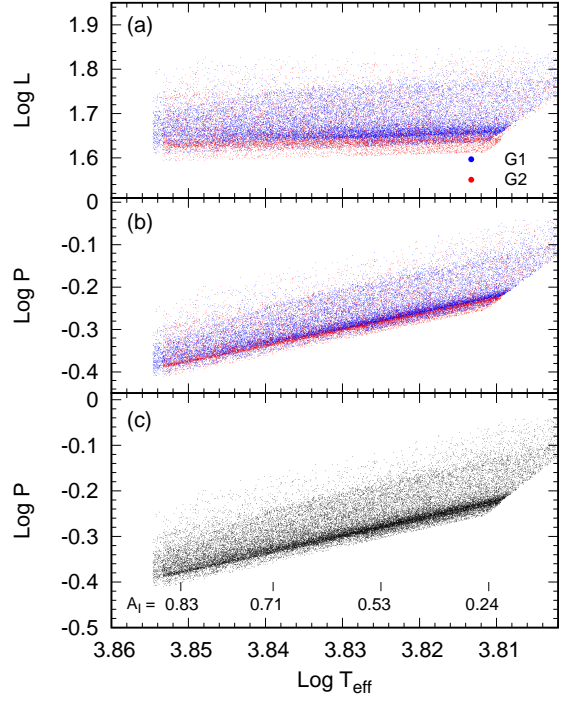}
\caption{Similar to Figure~\ref{fig3}, but for our models with $\Delta$[Fe/H] = 0.17 dex between G2 and G1. Note that these models with the metallicity difference cannot reproduce the observed two sequences of RR Lyrae stars in the bulge (see the text).\label{fig8}}
\end{figure}

\begin{figure}[ht!]
\vspace{10pt}
\figurenum{9}
\centering
\epsscale{1.15}
\plotone{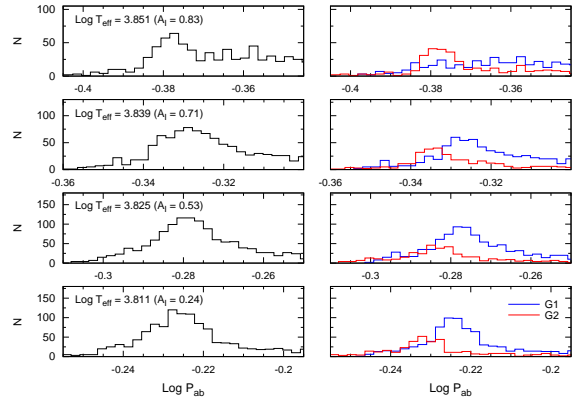}
\caption{Similar to Figure~\ref{fig4}, but for our models with $\Delta$[Fe/H] = 0.17 dex between G2 and G1. Note that the observed bimodal distribution is not reproduced in this case.\label{fig9}}
\end{figure}

As discussed above, despite the uncertainty, the metallicity difference between the two populations could be as large as 0.17 dex \citep{Pie15}. We have, therefore, explored in Figure~\ref{fig8} the effects of metallicity on the period-shift. For these models, we have adopted [Fe/H] = -1.1 for G1 and [Fe/H] = -0.93 for G2. Since the HB morphology is extremely sensitive to metallicity \citep{Lee94}, metal-rich G2 in this case are placed on the red HB unless they are assumed to be $\sim$1 Gyr older than metal-poor G1. Although this is highly unlikely in terms of chemical evolution, we have nevertheless adopted this ad-hoc assumption in order to place G2 in the IS. As is clear from Figure~\ref{fig8}, however, the observed split in period is not reproduced, and the histograms in Figure~\ref{fig9} just show broad monomodal distributions. Our additional simulations indicate that this conclusion is not affected by the assumed population ratio between G2 and G1. The absence of the split, in this case, is because the effect on period from the luminosity difference between G1 and G2 ($\Delta log$~L $\approx$ 0.02; see panel (a) of Figure~\ref{fig8}) is mostly cancelled out by the mass difference ($\Delta$M $\approx$ 0.02 $M_{\bigodot}$) at given temperature. This effect has been well known since the discovery of the Sandage period-shift effect among RR Lyrae stars in GCs \citep{San81,Swe87,Lee90,Jan14}, which suggests that the metallicity difference alone cannot produce the observed two sequences of RR Lyrae stars in the bulge. The same argument also implies that, even if we had included a small metallicity spread in the model simulations in Figure~\ref{fig3}, it would have only little impact on the result.\\

\section{A possible connection with the double red clump} \label{sec3}

\begin{figure}[ht!]
\figurenum{10}
\centering
\epsscale{1.95}
\plotone{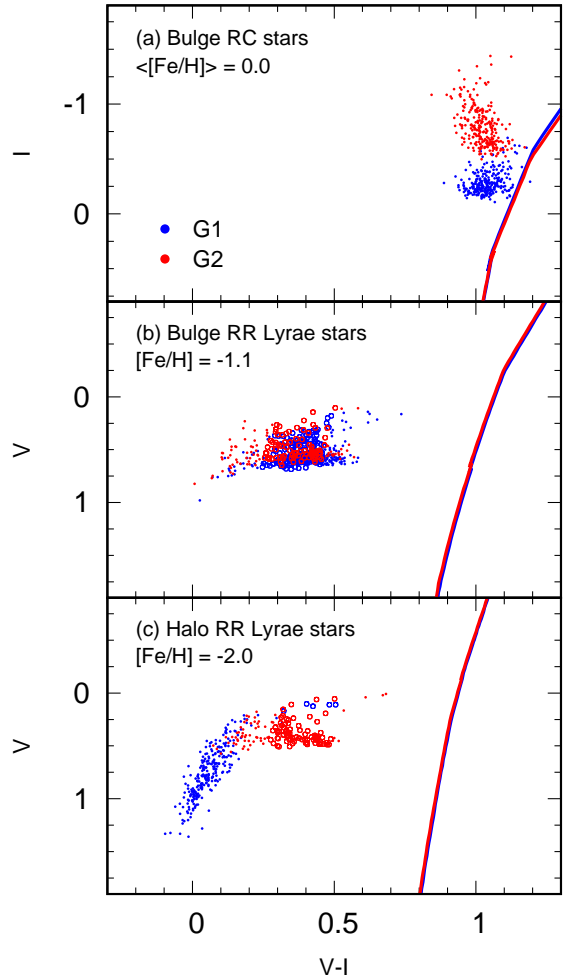}
\caption{Our synthetic HB models for G1 and G2 at different metallicity regimes.  Panel (a) is to reproduce the double RC in the metal-rich bulge as was suggested by Lee et al. (2015), while panel (b) is for the two populations of RR Lyrae stars in the relatively metal-poor bulge. Panel (c) is for the metal-poor Oosterhoff II GCs observed in the outer halo, following the models by \citet{Jan15}. Open circles are RR Lyrae variables.\label{fig10}}
\end{figure}

\begin{figure}[ht!]
\figurenum{11}
\centering
\epsscale{1.2}
\plotone{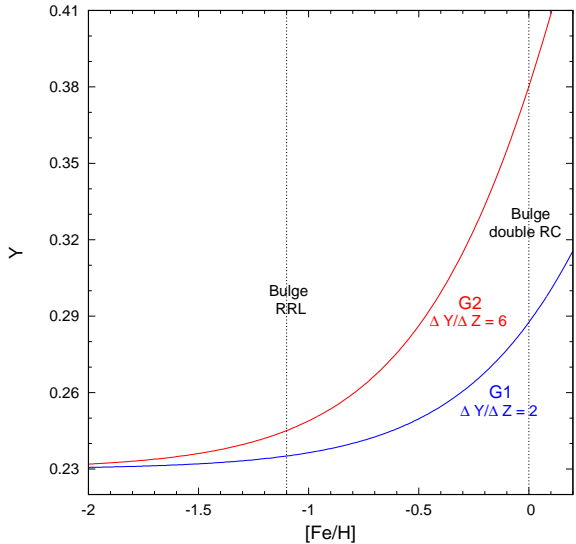}
\caption{Helium abundances for G1 and G2 as a function of [Fe/H]. G2 in our models follow the helium enrichment parameter $\Delta$Y/$\Delta$Z = 6, while G1 have normal helium abundance ($\Delta$Y/$\Delta$Z = 2). The difference in helium abundance between G2 and G1, therefore, strongly increases with metallicity.\label{fig11}}
\vspace{-7pt}
\end{figure}

As described above, the Oosterhoff dichotomy and the two populations of RR Lyrae stars observed in the halo and bulge could be naturally reproduced in the multiple population paradigm. That this effect might also reproduce the double red clump observed in the metal-rich regime of the bulge was recently suggested by \citet{Lee15}. Interestingly, we can infer from these studies for metal-poor and metal-rich populations that the helium abundance for G2 increases with metallicity and is well represented by the helium enrichment parameter $\Delta$Y/$\Delta$Z $\approx$ 6. In order to illustrate this, Figure~\ref{fig10} shows our population models for G1 and G2 on the HB and RC at different metallicity regimes. In these models, the helium abundance for G2 follows $\Delta$Y/$\Delta$Z $\approx$ 6, while G1 has normal helium abundance following $\Delta$Y/$\Delta$Z = 2,  as depicted in Figure~\ref{fig11}. Our models in panel (a) of Figure~\ref{fig10} are similar to those presented in \citet{Lee15} for the double RC in the bulge, and in panel (b) are identical to the bulge RR Lyrae models in Figure~\ref{fig2}. Table~\ref{tab1} lists parameters adopted in these models, where G2/G1 ratio is from \citet{Jan15} and \citet{Lee15} for the RR$_{\rm ab}$ stars and double RC respectively. Since G2 follows a fixed $\Delta$Y/$\Delta$Z = 6, the difference in helium abundance between G2 and G1, $\Delta$Y (G2 - G1), becomes an order of magnitude larger for the RC at solar metallicity compared to bulge RR Lyrae stars at [Fe/H] = -1.1. For comparison, models in panel (c) are for the typical metal-poor Oosterhoff II GCs in the outer halo as presented by \citet{Jan15}, where the age difference between G1 and G2 is larger ($\sim$~1.4 Gyr).

%\{From the WISE mid-IR image, \citet{Nes16} report a direct detection of faint X-shaped structure in the Milky Way bulge, but it is not clear from this work whether the stellar density in this faint structure is high enough to be observed as a strong double RC. Without critical distance information, therefore, the origin of the double RC is still an open question.}

\begin{figure}[ht!]
\figurenum{12}
\centering
\epsscale{1.17}
\plotone{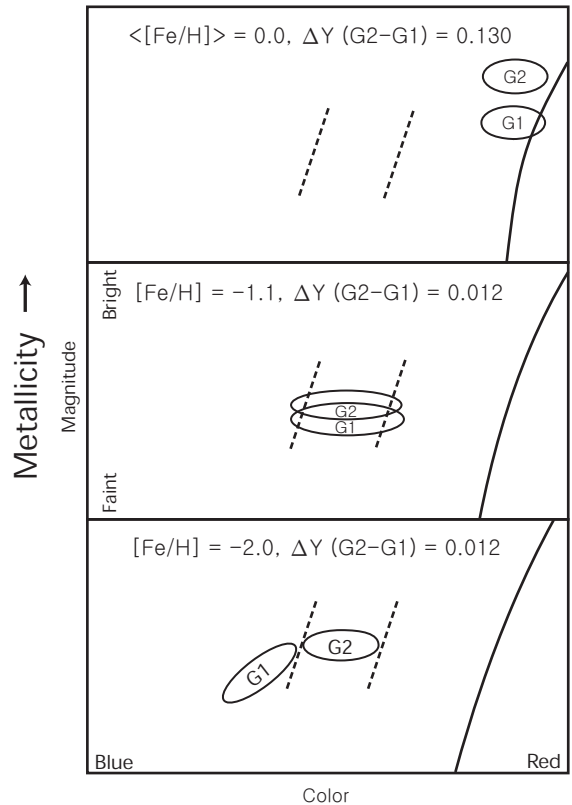}
\caption{Schematic diagrams (based on Figure~\ref{fig10}) illustrating different manifestations of the multiple population phenomenon at different metallicity regimes. In the most metal-poor regime (lower panel), the separation between G1 and G2 is large and therefore the IS is mostly populated by G2, representing Oosterhoff type II GCs in the outer halo. At intermediate metallicity (middle panel), the separation between G1 and G2 becomes smaller and both G1 and G2 are placed in the IS, producing two populations of RR Lyrae stars. In the metal-rich regime (upper panel), G1 and G2 are placed on the redder HB, showing a double red clump with a large difference in $\Delta$Y(G2 - G1).\label{fig12}}
\vspace{-1pt}
\end{figure}

Figure~\ref{fig12} is a schematic diagram, based on Figure~\ref{fig10}, which explains how the HB and RC features change in different metallicity regimes in our scenario. In the most metal-poor regime (lower panel), the separation between G1 and G2 is large and therefore the IS is mostly populated by G2, while G1 is placed on the blue HB. At intermediate metallicity (middle panel), the separation between G1 and G2 becomes smaller and both G1 and G2 are placed in the IS, producing two populations of RR Lyrae stars as described in Section \ref{sec2}. Note that in these relatively metal-poor regimes, $\Delta$Y(G2 - G1) is small ($\sim$0.012).  In the metal-rich regime (upper panel) however, this difference, following $\Delta$Y/$\Delta$Z = 6 for G2, becomes significantly large ($\Delta$Y $\approx$ 0.13), and G1 and G2 are placed on the redder HB, showing a double red clump with larger luminosity difference. In the metallicity regime between the upper and middle panels, it is encouraging to confirm that RC models for G1 and G2 in NGC 6441 ([Fe/H] $\approx$ -0.5, $\Delta$Y $\approx$ 0.05; \citealt{Cal07}) and Terzan 5 ([Fe/H] $\approx$ -0.2, $\Delta$Y $\approx$ 0.09; \citealt{Lee15}) have intermediate values for $\Delta$Y(G2 - G1), following the trend in Figure~\ref{fig11}. Therefore, it is most likely from our models that the Oosterhoff period groups, two populations of RR Lyrae stars, and the double RC observed in the halo and bulge respectively, are other manifestations of the same multiple population phenomenon observed in halo globular clusters.\\

\section{Discussion} \label{sec4}

The main hypothesis in our models is that the helium abundance of G2 increases strongly with metallicity following a higher helium enrichment parameter $\Delta$Y/$\Delta$Z = 6. How could this be explained in terms of chemical evolution? In our scenario, most stars in a classical bulge component would have been provided by disruption of primordial building blocks in a hierarchical merging paradigm. It is assumed that pristine gas in these building blocks was globally enriched by supernovae following the standard helium enrichment parameter ($\Delta$Y/$\Delta$Z $\approx$ 2) before the formation of G1, while the gas that formed G2 is likely to be locally enriched within these systems by helium-rich winds of first-generation massive stars. Most of the supernova ejecta would have escaped from these relatively less massive systems with little effect on the pre-enriched gas retained in them \citep[see, e.g.,][]{Rom10,Ten15}. In such a case, our chemical evolution models (J. Kim \& Y.-W. Lee 2016, in prep.) confirm that the trend for G2 in Figure~\ref{fig11} is indeed predicted from the chemical enrichments by the winds of massive stars, together with the winds and ejecta from low and intermediate mass asymptotic-giant-branch stars. This is mainly due to the strong metallicity dependence of helium yield from the winds of massive stars \citep[see][]{Mae92,Mey08}. The same effect can also explain why the G2/G1 ratio apparently increases with metallicity (see Table~\ref{tab1}).

%Our results for the bulge RR Lyrae and RC stars suggest rather extended star formation time scale in the classical bulge. RR Lyrae stars represent metal-poor ([Fe/H] $\approx$ -1.1) population in the bulge, while RC stars are mostly from metal-rich ([Fe/H] $\approx$ 0.0) population.  According to Table 1, the age of G1 for RR Lyrae population is $\sim$13.5 Gyr, while that for G2 of RC stars is about $\sim$9 Gyr. Therefore, the age difference between the oldest (G1 of RR$_{\rm ab}$) and youngest (G2 of RC) populations is predicted to be about $\sim$4.5 Gyr. Since the RC feature does not change radically at 3 - 4 Gyr younger ages (S.-J. Joo et al., in prep) from those listed in Table 1, this age difference should be considered as a lower limit. This result would be affected if the metallicity dependence of the mass-loss on the red giant-branch does not follow the trend predicted from the Reimers (1977) formula. Nevertheless, this age difference appears to be roughly consistent with that inferred from the main sequence turn-off age datings from GCs and bulge field (Clarkson et al. 2008; Dotter et al. 2010; Valenti et al. 2013; Bensby et al. 2013). 

As described in Section \ref{sec2}, our models for the bulge RR Lyrae stars are almost identical to those for the inner halo GCs with similar metallicity. This suggests that the same split in the amplitude-period diagram would have been observed in these GCs if they had contained a sufficiently large sample of RR$_{\rm ab}$ stars. Similarly, the Oosterhoff dichotomy is not only observed in GCs but also in halo fields \citep{Lee90a,Ses13}. Therefore, the fact that two populations of stars are present in the classical bulge and halo fields, similarly to the case of GCs, would suggest then that GCs are remaining relics of the building blocks that provided the bulge and halo fields with two populations of stars originating from G1 and G2. It is important to note that G2 in our definition experienced just a modest helium enhancement ($\Delta$Y/$\Delta$Z $\approx$ 6), and therefore they appear to be different from third generation stars (G3) observed in GCs which require very extreme helium enrichment parameter ($\Delta$Y/$\Delta$Z $>$ 70) and special formation condition in the central region of a proto-GC \citep[see, e.g.,][]{Der08}. Our models presented in this paper suggest that the presence of G1 and G2 with different helium abundances is a ubiquitous phenomenon throughout the Milky Way halo and classical bulge. Sooner or later, the direct trigonometric parallax distance measurements from Gaia will show whether this phenomenon is indisputably extended to the double RC in the metal-rich regime of the bulge.\\

We thank the referee for a number of helpful suggestions.
Support for this work was provided by the National Research Foundation of Korea to the Center for Galaxy Evolution Research.




\begin{thebibliography}{}
%\bibitem[\protect\citeauthoryear{Bono et al.}{1997}]{Bon97} Bono, G., Caputo, F., Castellani, V., \& Marconi, M. 1997, A\&AS, 121, 327

\bibitem[\protect\citeauthoryear{Caloi \& D'Antona}{2007}]{Cal07} Caloi, V., \& D'Antona, F. 2007, A\&A, 463, 949

\bibitem[\protect\citeauthoryear{Carretta et al.}{2009}]{Car09} Carretta, E., Bragaglia, A., Gratton, R.~G., et al. 2009, A\&A, 505, 117
\bibitem[\protect\citeauthoryear{Corwin \& Carney}{2001}]{Cor01} Corwin, T. M., \& Carney, B. W. 2001, AJ, 122, 3183

\bibitem[\protect\citeauthoryear{D'Antona \& Caloi}{2004}]{Dan04} D'Antona, F., \& Caloi, V. 2004, ApJ, 611, 871

\bibitem[\protect\citeauthoryear{D'Ercole et al.}{2008}]{Der08} D'Ercole, A., Vesperini, E., D'Antona, F., McMillan, Stephen, L.~W., \& Recchi, S. 2008, MNRAS, 391, 825

\bibitem[\protect\citeauthoryear{Gonzalez et al.}{2015}]{Gon15} Gonzalez, O. A., Zoccali, M., Debattista, V. P., et al. 2015, A\&A, 583, L5

\bibitem[\protect\citeauthoryear{Gratton et al.}{2012}]{Gra12} Gratton, R.~G., Carretta, E., \& Bragaglia, A. 2012a, A\&A Rev., 20, 50

\bibitem[\protect\citeauthoryear{Han et al.}{2009}]{Han09} Han, S.-I., Kim, Y.-C., Lee, Y.-W., et al. 2009, Globular Clusters - Guides to Galaxies, Springer, Berlin, p. 33

\bibitem[\protect\citeauthoryear{Jang et al.}{2014}]{Jan14} Jang, S., Lee, Y.-W., Joo, S.-J., \& Na, C. 2014, MNRAS, 443, L15

\bibitem[\protect\citeauthoryear{Jang \& Lee}{2015}]{Jan15} Jang, S., \& Lee, Y.-W. 2015, ApJS, 218, 31

\bibitem[\protect\citeauthoryear{Joo \& Lee}{2013}]{Joo13} Joo, S.-J., \& Lee, Y.-W. 2013, ApJ, 762, 36

\bibitem[\protect\citeauthoryear{Joo et al.}{2016}]{Joo16} Joo, S.-J., Lee, Y.-W., \& Chung, C. 2016, arXiv:1609.01294

\bibitem[\protect\citeauthoryear{Kunder \& Chaboyer}{2008}]{Kun08} Kunder, A., \& Chaboyer, B. 2008, 136, 2441K 


%\bibitem[\protect\citeauthoryear{Kunder et al.}{2013}]{Kun13} Kunder, A., Stetson, P. B., Cassisi, S., et al. 2013, AJ, 146, 119


\bibitem[\protect\citeauthoryear{Lee}{1990}]{Lee90a} Lee, Y.-W. 1990, ApJ, 363, 159

\bibitem[\protect\citeauthoryear{Lee}{1992}]{Lee92} Lee, Y.-W. 1992, AJ, 104, 1780

\bibitem[\protect\citeauthoryear{Lee et al.}{1990}]{Lee90} Lee, Y.-W., Demarque, P., \& Zinn, R. 1990, ApJ, 350, 155

\bibitem[\protect\citeauthoryear{Lee et al.}{1994}]{Lee94} Lee, Y.-W., Demarque, P., \& Zinn, R. 1994, ApJ, 423, 248

\bibitem[\protect\citeauthoryear{Lee et al.}{2015}]{Lee15} Lee, Y.-W., Joo, S.-J., \& Chung, C. 2015, MNRAS, 453, 3906

\bibitem[\protect\citeauthoryear{Lee et al.}{2005}]{Lee05} Lee, Y.-W., Joo, S.-J., Han, S.-I., et al. 2005, ApJL, 621, L57

\bibitem[\protect\citeauthoryear{Lopez-Corredoira}{2016}]{Lop16} Lopez-Corredoira,~M. 2016, arXiv:1606.09627

\bibitem[\protect\citeauthoryear{McWilliam \& Zoccali}{2010}]{Mcw10} McWilliam, A. \& Zoccali, M. 2010, ApJ, 724, 1491

\bibitem[\protect\citeauthoryear{Maeder}{1992}]{Mae92} Maeder, A. 1992, A\&A, 264,105M

\bibitem[\protect\citeauthoryear{Meynet}{2008}]{Mey08} Meynet, G. 2008, EAS, 32, 187M 

\bibitem[\protect\citeauthoryear{Nataf et al.}{2010}]{Nat10} Nataf, D. M., Udalski, A., Gould, A., Fouqu\'{e}, P., \& Stanek, K. Z. 2010, ApJ, 721, 28

\bibitem[\protect\citeauthoryear{Ness \& Lang}{2016}]{Nes16} Ness, M., \& Lang, D. 2016, AJ,152,14

\bibitem[\protect\citeauthoryear{Pietrukowicz et al.}{2015}]{Pie15} Pietrukowicz, P., Koz$\l$owski, S., Skowron, J., et al. 2015, ApJ, 811, 113

\bibitem[\protect\citeauthoryear{Reimers}{1977}]{Rei77} Reimers, D. 1977, A\&A, 57, 395
\bibitem[\protect\citeauthoryear{Romano et al.}{2010}]{Rom10} Romano, D., Karakas, A. I., Tosi, M., \& Matteucci, F. 2010, A\&A, 552, 32
\bibitem[\protect\citeauthoryear{Sandage et al.}{1981}]{San81} Sandage, A., Katem, B., \& Sandage, M. 1981, ApJS, 46, 41
\bibitem[\protect\citeauthoryear{Sesar et al.}{2013}]{Ses13} Sesar, B., Ivezi\'{c}, \v{Z}., Stuart, J.~S. et al. 2013, AJ, 146, 21
\bibitem[\protect\citeauthoryear{Soszy\'nski et al.}{2014}]{Sos14} Soszy\'nski, I., Udalski, A., Szyma\'nski, M. K., et al. 2014, AcA, 64, 177
\bibitem[\protect\citeauthoryear{Sweigart et al.}{1987}]{Swe87} Sweigart, A. V., Renzini, A., \& Tornambe, A. 1987, ApJ, 312, 762

\bibitem[\protect\citeauthoryear{Tenorio-Tagle et al.}{2015}]{Ten15} Tenorio-Tagle, G., Mu\~{n}oz-Tu\~{n}\'{o}n, C., Silich, S., \& Cassisi, S. 2015, ApJ, 814, 8


\bibitem[\protect\citeauthoryear{VandenBerg et al.}{2012}]{Van12} VandenBerg, D. A., Bergbusch, P. A., Dotter, A., et al. 2012, ApJ, 755, 15


\bibitem[\protect\citeauthoryear{Walker \& Terndrup}{1991}]{Wal91} Walker, A. R., \& Terndrup, D. M. 1991, ApJ, 378, 119W


\bibitem[\protect\citeauthoryear{Yong et al.}{2015}]{Yon15} Yong, D., Grundahl, F., \& Norris, J. E. 2015, MNRAS, 446, 3319


\end{thebibliography}
\end{document}